\documentclass[12pt]{article}

\usepackage{amsmath,amssymb}
\usepackage{graphicx}

\makeatletter

\renewcommand\section{\@startsection{section}{1}{\z@}
                                   {-3.5ex \@plus -1ex \@minus -.2ex}
                                   {2.3ex \@plus .2ex}
                                   {\normalfont\large\bfseries}}
\renewcommand\subsection{\@startsection{subsection}{2}{\z@}
                                   {-3.25ex\@plus -1ex \@minus -.2ex}
                                   {1.5ex \@plus .2ex}
                                   {\normalfont\normalsize\bfseries}}
\renewcommand\subsubsection{\@startsection{subsubsection}{3}{\z@}
                                   {-3.25ex\@plus -1ex \@minus -.2ex}
                                   {1.5ex \@plus .2ex}
                                   {\normalfont\normalsize\bfseries}}
\renewcommand\paragraph{\@startsection{paragraph}{4}{\z@}
                                   {3.25ex \@plus1ex \@minus.2ex}
                                   {-1em}
                                   {\normalfont\normalsize\bfseries}}

\makeatother

\newcommand{\beq}{\begin{equation}}
\newcommand{\eeq}{\end{equation}}
\newcommand{\bea}{\begin{eqnarray}}
\newcommand{\eea}{\end{eqnarray}}

\newcommand{\SL}{{\rm SL}}
\newcommand{\SU}{{\rm SU}}
\newcommand{\SO}{{\rm SO}}
\newcommand{\Sp}{{\rm Sp}}
\newcommand{\Spin}{\rm Spin}

\newcommand{\U}{{\rm U}}

\newcommand{\C}{\mathbb C}
\newcommand{\Q}{\mathbb Q}
\newcommand{\R}{\mathbb R}
\newcommand{\Z}{\mathbb Z}

\newcommand{\id}{\hbox{1\kern-.27em l}}

\newcommand{\cA}{{\cal A}}

\newcommand{\cG}{{\cal G}}
\newcommand{\cH}{{\cal H}}

\newcommand{\cJ}{{\cal J}}

\newcommand{\cL}{{\cal L}}

\newcommand{\cR}{{\cal R}}
\newcommand{\cS}{{\cal S}}

\newcommand{\cU}{{\cal U}}

\begin{document}

\pagestyle{empty}

\begin{center}

\vspace*{30mm}
{\Large  The partition bundle of type $A_{N - 1}$ $(2, 0)$ theory}

\vspace*{30mm}
{\large M{\aa}ns Henningson}

\vspace*{5mm}
Department of Fundamental Physics\\
Chalmers University of Technology\\
S-412 96 G\"oteborg, Sweden\\[3mm]
{\tt mans@chalmers.se}     
     
\vspace*{30mm}{\bf Abstract:} 
\end{center}
Six-dimensional $(2, 0)$ theory can be defined on a large class of six-manifolds endowed with some additional topological and geometric data (i.e. an orientation, a spin structure, a conformal structure, and an $R$-symmetry bundle with connection). We discuss the nature of the object that generalizes the partition function of a more conventional quantum theory. This object takes its values in a certain complex vector space, which fits together into the total space of a complex vector bundle (the `partition bundle') as the data on the six-manifold is varied in its infinite-dimensional parameter space. In this context, an important role is played by the middle-dimensional intermediate Jacobian of the six-manifold endowed with some additional data (i.e. a symplectic structure, a quadratic form, and a complex structure). We define a certain hermitian vector bundle over this finite-dimensional parameter space. The partition bundle is then given by the pullback of the latter bundle by the map from the parameter space related to the six-manifold to the parameter space related to the intermediate Jacobian.

\newpage \pagestyle{plain}

\section{Introduction}
Six-dimensional $(2, 0)$ theory is a comparatively new kind of quantum theory, which in many respects is rather different from how we think of e.g. quantum field theory \cite{Witten95}. These theories are remarkably unique, and can be completely specified by the type
\bea
\Phi & \in & \mathrm{ADE} \cr
& \simeq & \{ \mathrm{simply \; laced \; Lie \; algebras} \} \cr
& \simeq & \{ \mathrm{finite \; subgroups \; of \;} \SU (2) \} .
\eea

A $(2, 0)$ theory can be defined on an arbitrary six-manifold $M$ which is orientable and admits a spin structure, i.e. the first two Stiefel-Whitney classes of its tangent bundle must vanish: 
\beq
w_1 (TM) = w_2(TM) = 0 . 
\eeq
Furthermore, $M$ must be endowed with some additional topological and geometrical data:
\bea
\sigma & \in & \Sigma \cr
& = & \{ \mathrm{orientations \; on \;} M \} \cr
& = & \mathrm{affine \; space \; over \;} H^0 (M, \Z_2) \cr
s & \in & \cS \cr
& = & \{ \mathrm{spin \; structures \; on \;} M \} \cr
& = & \mathrm{affine \; space \; over \;} H^1 (M, \Z_2) \cr
[g] & \in & \cG \cr
& = & \{ \mathrm{conformal \; structures \; on \;} M \} \cr
& = & \mathrm{infinite \; dimensional \; real \; manifold}
\eea
as well as data related to the $R$-symmetry:
\bea
R & \in & \cR \cr
& = & \{ \mathrm{principal \;} \Sp (4) \simeq \Spin (5) \mathrm{\; bundles \; over \;} M \} \cr
& = & \mathrm{discrete \; set \; labeled \; by \; characteristic \; classes \; in \;} H^4 (M, \Q) \mathrm{\; and \;} H^5 (M, \Z_2) \cr
A & \in & \cA \cr
& = & \{ \mathrm{connections \; on \;} R \} \cr
& = & \mathrm{affine \; space \; over \;} \Omega^1 (M, \mathrm{ad} (R)) .
\eea
Our main concern in this paper is to investigate, for a fixed type $\Phi$ and a fixed topological class of $M$, the dependence of the theory on the data $(\sigma, s, [g])$ and $(R, A)$. In particular, we will try to elucidate the nature of an object $Z$ that generalizes the partition function of a more conventional quantum theory.

A first remark is that, because of various `anomalies', there is a phase ambiguity in $Z$:
\begin{itemize}
\item
The conformal anomaly \cite{Henningson-Skenderis} implies a dependence on the choice of representative metric $g$ for the conformal structure $[g] \in \cG$.
\item
The chiral, gravitational and mixed chiral/gravitational anomalies \cite{Harvey-Minasian-Moore, Intriligator} imply a dependence on the choice of trivialization of the $\Sp (4)$ bundle $R$ and the parametrization of $M$.
\end{itemize}
These phenomena clearly provide extremely important clues for a future complete definition of $(2, 0)$ theory, but will nevertheless be disregarded in this paper. Our ambition here is thus only to describe the nature of $Z$ up to a complex `anomalous' phase factor. (Even so, we are not attempting to actually determine $Z$, although we hope that this will eventually be possible; in this paper we will just answer the question of what kind of object $Z$ is.)

For simplicity, we will restrict ourselves to the case where $R$ is a trivial bundle and $A$ is the trivial connection. Furthermore, we will only consider $(2, 0)$ theory of type $\Phi = A_{N - 1}$ for $N = 2, 3, \ldots$, since this allows us to use results from holography \cite{Maldacena}. But we hope to continue to more general cases in the near future.

The next important point is that, for fixed data $(\sigma, s, [g])$, $Z$ is not a single complex number, but rather an element of a certain finite-dimensional complex vector space $V$ \cite{Witten98}. (This should not be confused with the more familiar quantum mechanical Hilbert space $\cH$, that appears when the theory is considered on a manifold $M$ with a prefered (Euclidean) `time' direction). Indeed, the data $(\sigma, s, [g])$ on $M$ determine data related to the intermediate Jacobian
\beq
T = H^3 (M, \R) / H^3 (M, \Z) ,
\eeq
which is a torus of dimension 
\beq
\dim_\R T = 2 n = b_3 (M). 
\eeq
These data are:
\bea
\omega & \in & \Omega \cr
& = & \{ \mathrm{symplectic \; structures \; on \;} H^3 (M, \R) \mathrm{\; induced \; from \; the \; intersection \; form} \} \cr
& = & \mathrm{set \; with \; 2 \; elements} \cr
u & \in & \cU \cr
& = & \{ \mathrm{non{-}degenerate \; quadratic \; forms \; on \;} H^3 (M, \Z_2) \mathrm{\; polarized \; by \; \omega} \} \cr
& = & \mathrm{set \; with \; 2^{2 n} \; elements} \cr
J & \in & \cJ \cr
& = & \{ \mathrm{translation \; invariant \; complex \; structures \;  on \;} T \} \cr
& = & \mathrm{complex \; space \; of \; dimension \;} \frac{1}{2} n (n + 1) .
\eea
The data $(\omega, u, J)$ define a hermitian line-bundle $\cL$ over $T$ \cite{Witten96, Belov-Moore}, and for $(2, 0)$ theory of type $\Phi = A_{N - 1}$, the complex vector space $V$ in which $Z$ takes its values can be identified with the space $H^0 (T, \cL^N)$ of holomorphic sections of $\cL^N$ \cite{Witten98}. These matters are reviewed in the next section. 

In section three, we describe how the family of vector spaces $V$ fit together into the total space of a complex vector bundle over the parameter space $\Sigma \times \cS \times \cG$ as the data $(\sigma, s, [g])$ are varied. $Z$ should thus be understood as the `partition section' of this `partition bundle'. (This is largely implicit in earlier work on $(2, 0)$ theory \cite{Witten98, Witten09}; our aim here is merely to make these results somewhat more explicit.) In fact, the family of spaces $H^0 (T, \cL^N)$ naturally fit together to the total space of a hermitian vector bundle over the finite-dimensional parameter space $\Omega \times \cU \times \cJ$ in which the data $(\omega, u, J)$ take its values. More precisely, for fixed values of $\omega \in \Omega$ and $u \in \cU$, we can define a hermitian vector bundle $\tilde{E}$ of rank $N^n$ over $\overline{\cJ} / \Lambda_{(\omega, u)}$, where $\overline{\cJ}$ denotes the universal covering space of $\cJ$, and $\Lambda_{(\omega, u)}$ is the subgroup of the automorphism group of $H^3 (M, \Z)$ that leaves these data invariant. We have a map
\beq
\phi \colon \overline{\cG} / \Lambda_{(\sigma, s)} \rightarrow \overline{\cJ} / \Lambda_{(\omega, u)}
\eeq
which determines the complex structure $J$ on $T$ in terms of the conformal structure $[g]$ on $M$. Here $\overline{\cG}$ is the universal covering space of $\cG$, and $\Lambda_{(\sigma, s)}$ is the subgroup of the mapping class group of $M$ that stabilizes the data  $\sigma \in \Sigma$ and $s \in \cS$. The partition bundle $E$ is now the complex vector bundle over the infinite dimensional space $\overline{\cG} / \Lambda_{(\sigma, s)}$ given by the pullback of $\tilde{E}$ by the map $\phi$:
\beq
E = \phi^* (\tilde{E}) .
\eeq

In section four, we make these somewhat abstract considerations more concrete by choosing a parametrization of $\overline{\cJ}$ so that it can be identified with a Siegel generalized upper half-space. This allows us to determine the transition functions of the hermitian vector bundle $\tilde{E}$ over the moduli space $\overline{\cJ} / \Lambda_{(\omega, u)}$. We find that these are related to a Siegel modular form and a flat vector bundle.

In an appendix, we exemplify the determination of the quadratic form $u$ for all possible choices of spin structure $s \in \cS$ in the particular case of $M = T^6$.

\section{The vector space}
This section is much influenced by \cite{Witten96, Witten98}.

The symplectic structure $\omega \in \Omega^2 (T)$ on the middle dimensional intermediate Jacobian $T = H^3 (M, \R) / H^3 (M, \Z)$ of $M$ only depends on the orientation $\sigma \in \Sigma$ on $M$ and is given by
\beq
\omega [\delta_1 C, \delta_2 C] = \int_M \delta_1 C \wedge \delta_2 C .
\eeq
Here $\delta_1 C, \delta_2 C \in H^3 (M, \R)$ are regarded as tangent vectors to $T$. Clearly, $\omega$ is closed (in fact constant) and
\beq
\int_{T} \frac{1}{n !} \omega^n = 1 ,
\eeq
so the de Rham cohomology class $[\omega] \in H^2_{\rm de \, Rham} (T)$ lies in the image of the inclusion $H^2 (T, \Z) \subset H^2 (T, \R) \simeq H^2_{\rm de \, Rham} (T)$.

By the de Rham theorem and the Hodge theorem
\beq
H^3 (M, \R) \simeq H^3_\mathrm{de \, Rham} (M) \simeq \Omega^3_\mathrm{harmonic} (M) ,
\eeq
where $\Omega^3_\mathrm{harmonic} (M)$ are the three-forms on $M$ that are harmonic with respect to the conformal structure $[g]$ on $M$. Together with the orientation $\sigma \in \Sigma$ on $M$, $[g] \in \cG$ determines the Hodge duality operator
\beq
* \colon \Omega^3_\mathrm{harmonic} (M) \rightarrow \Omega^3_\mathrm{harmonic} (M) .
\eeq
This operator obeys $* * = - 1$, and thus defines a complex structure $J \in \cJ$ on $T$, i.e. we have (for given $\sigma \in \Sigma$) a map
\beq
\phi \colon \cG \rightarrow \cJ .
\eeq
The symplectic structure $\omega \in \Omega^2 (T)$ is of type $(1, 1)$ with respect to the conformal structures defined in this way, and thus gives $T$ the structure of a (flat) K\"ahler manifold.

We will now explain how the spin structure $s \in \cS$ on $M$ determines a quadratic form
\beq
u \colon H^3 (M, \Z_2) \rightarrow \Z_2 .
\eeq
A straight line from $0$ to an element $\gamma \in H^3 (M, \Z)$ descends to a closed curve on $T$, and thus determines a three-form gauge-field $C \in \Omega^3 (S^1 \times M)$ with field-strength $G = d C$. We let $X$ be an eight-dimensional open spin manifold, which bounds $S^1 \times M$ and over which the product of the anti-periodic spin structure on $S^1$ and the given spin structure on $M$ extends. (By a result in cobordism theory, such a manifold always exists.) Extending $G$ to $X$ we then define $u (\gamma) \in \Z_2$ by
\beq
(-1)^{u (\gamma)} = \exp \left( 2 \pi i \frac{1}{2} \int_X (G \wedge G - \frac{p_1}{2} \wedge \frac{p_1}{2}) \right) ,
\eeq
where $p_1$ denotes the first Pontryagin class of $T X$ (which can be divided by two in a canonical way for $X$ spin). Clearly, this expression only depends on the reduction of $\gamma$ modulo two. Furthermore, it is in fact well-defined and independent of the choice of $X$ despite the half-integer prefactor \cite{Dijkgraaf-Witten}. Finally, it obeys the condition
\beq \label{curvature}
u (\gamma) + u (\gamma^\prime) = \omega[\gamma, \gamma^\prime] + u (\gamma + \gamma^\prime) ,
\eeq
for $\gamma, \gamma^\prime \in H^3 (M, \Z_2)$, i.e. the symplectic structure $\omega$ acts as a polarization of $u$.  

The data $(\omega, u, J)$ determine a hermitian line bundle $\cL$ over $T$: Its curvature equals the symplectic structure $\omega$, and the holonomy along the closed curve defined by a straight line from $0$ to $\gamma \in H^3 (M, \Z)$ is given by the quadratic form $u$ as $(-1)^{u (\gamma)}$. We will be mostly interested in the $N$th power $\cL^N$ of $\cL$. This bundle has curvature $N \omega$. The holonomies along the curves $\gamma$ described above are all trivial for even $N$ (in which case $\cL^N$ is actually independent of the spin structure $s \in \cS$), and agree with the holonomies of $\cL$ for odd $N$.

For $(2, 0)$ theory of type $\Phi = A_{N - 1}$, the vector space $V$ in which $Z$ takes its values can now be determined by geometric quantization as the Hilbert space of a certain topological field theory of Schwarz type \cite{Schwarz}. This field theory governs the three-form gauge field $C$ that is part of the holographic gravity dual of the $(2, 0)$ theory on an open seven manifold $Y$ that bounds $M$. In this way, one finds that $V$ is given by the space of holomorphic sections of the bundle $\cL^N$:
\beq
V \simeq H^0 (T, \cL^N) .
\eeq
The Hilbert space inner product on $V$ is
\beq
\langle s | s^\prime \rangle = \int_{T} \frac{1}{n !} \omega^n \, (s, s^\prime) 
\eeq
for $s, s^\prime \in H^0 (T, \cL^N)$, where $( . , . )$ is the Hermitian structure on the fibers of $\cL$.

To avoid a possible misunderstanding, we point out that this does not mean that the $(2, 0)$ theory on $M = \partial Y$ couples to a three-form gauge field $C$. Indeed, in a situation with a stack of $N$ parallel $M5$-branes in $M$-theory, the $C$-field of $M$-theory couples to the collective `center of mass' degrees of freedom of the branes rather than to the $(2, 0)$ theory of type $\Phi = A_{N - 1}$ defined on the world-volume of the branes. This is analogous to a situation with a stack of $N$ parallel $D3$-branes in type IIB string theory, where the Yang-Mills theory on the branes has gauge group $\SU (N)$ rather than $\U (N) \simeq (\SU (N) \times \U (1)) / \sim$. (The quotient is by the equivalence relation $(\eta \id_N, \eta^{-1}) \sim (\id_N, 1)$ for $\eta$ an $N$th root of unity.) Here it is the $U (1)$ factor that represents the collective degrees of freedom.

Returning to the vector space $V \simeq H^0 (T, \cL^N)$, it follows from the Kodaira vanishing theorem, that the higher cohomology groups $H^k (T, \cL^N)$ for $1 \leq k \leq n = \dim_\C T$ vanish. The Hirzebruch-Riemann-Roch formula then gives
\bea
\dim V & = & \sum_{k = 0}^n (-1)^k \dim H^k (T, \cL^N) \cr
& = & \int_{T} e^{c_1 (\cL^N)} \mathrm{Td} (T) \cr
& = & \int_{T} \frac{1}{n !} (N \omega)^n \cr
& = & N^n ,
\eea
where we have used that the total Todd character $\mathrm{Td} (T) = 1$ for the flat manifold $T$ and first the Chern class $c_1 (\cL^N) = N [\omega]$. 

The bundle $\cL$ over $T = H^3 (M, \R) / H^3 (M, \Z)$ has no non-trivial isometries, but $\cL^N$ is invariant under translations 
\beq
T_c \colon T \rightarrow T
\eeq
by elements $c \in \frac{1}{N} H^3 (M, \Z) \subset H^3 (M, \R)$. Obviously  $T_c^N = \mathrm{id}$ and $T_c T_{c^\prime} = T_{c^\prime} T_c$. But the induced pullback maps
\beq
T_c^* \colon H^0 (T, \cL^N) \rightarrow H^0 (T, \cL^N)
\eeq
instead obey the relations
\beq
(T_c^*)^N = (-1)^{u (N c)} 
\eeq
and the Heisenberg relations
\beq
T_c^* T_{c^\prime}^* = T_{c^\prime}^* T_c^* \exp \left(N c \wedge c^\prime \right) .
\eeq
In the last formula, we have used the convenient short-hand notation
\beq
\exp (v) = \exp \left( 2 \pi i \int_M v \right)
\eeq
for $v \in H^6 (M, \C)$.

\section{The vector bundle}
There is a natural homomorphism from the mapping class group of $M$ to the subgroup, isomorphic to $\Sp_{2 n} (\Z)$, of the automorphism group of $H^3 (M, \Z)$ that stabilizes the symplectic structure $\omega$. This symplectic subgroup permutes the possible quadratic forms $u$ fulfilling (\ref{curvature}) via its reduction modulo two, which is isomorphic to the group $\Sp_{2 n} (\Z_2)$ of order 
\beq
| \Sp_{2 n} (\Z_2) | = 2^{n^2} \prod_{i = 1}^n (2^{2 i} - 1) .
\eeq
There are two orbits under this action:

The first orbit consist of quadratic forms $u$ for which $H^3 (M, \Z)$ admits a decomposition
\beq \label{AB-decomposition}
H^3 (M, \Z) = A \oplus B
\eeq
such that 
\beq \label{Lagrangian}
\int_M n \wedge n^\prime = \int_M m \wedge m^\prime = 0
\eeq
for $n, n^\prime \in A$ and $m, m^\prime \in B$ and furthermore
\beq \label{hyperbolic}
u (a + b) = \int_M a \wedge b
\eeq
for $a \in A \otimes \Z_2$ and $b \in B \otimes Z_2$. This means that $u$ gives $H^3 (M, \Z_2)$ the structure of the direct sum of $n$ hyperbolic planes. We have
\beq \label{distribution}
u (\gamma) = \left\{ \begin{array}{lll}
0 & \mathrm{\; for \;} 2^{2 n - 1} + 2^n - 2^{n - 1} & \mathrm{\; values \; of \;} \gamma \in H^3 (M, \Z_2) \cr
1 & \mathrm{\; for \;} 2^{2 n - 1} - 2^{n - 1} & \mathrm{\; values \; of \;} \gamma \in H^3 (M, \Z_2) .
\end{array} \right.
\eeq
(The sum of these numbers of course equals the cardinality $2^{2 n}$ of $H^3 (M, \Z_2)$.) The stabilizer in $\Sp_{2 n} (\Z_2)$ of such a quadratic form $u$ is isomorphic to the group ${\rm O}^+_{2 n} (\Z_2)$ of order
\beq
| {\rm O}^+_{2 n} (\Z_2) | = 2 \cdot 2^{n (n - 1)} (2^n - 1) \prod_{i = 1}^{n - 1} (2^{2 i} - 1) ,
\eeq
so the cardinality of the orbit is
\beq
\frac{| \Sp_{2 n} (\Z_2) |}{| {\rm O}^+_{2 n} (\Z_2) |} = 2^{2 n - 1} + 2^{n -1} .
\eeq

The second orbit consist of quadratic forms $u$ for which (\ref{hyperbolic}) is replaced by
\beq \label{anisotropic}
u (a + b) = \int_M a \wedge b + (\int_M a \wedge b_s)^2 + (\int_M a_s \wedge b)^2
\eeq
for some non-zero $a_s \in A \otimes \Z_2$ and $b_s \in B \otimes \Z_2$. $H^3 (M, \Z_2)$ now has the structure of the direct sum of $n - 1$ hyperbolic planes and a two-dimensional anisotropic space. There number of values of $\gamma \in H^3 (M, \Z_2)$ for which $u (\gamma) = 0$ or $u (\gamma) = 1$ are interchanged as compared to the quadratic forms on the first orbit. The stabilizer of $u$ is isomorphic to the group ${\rm O}^-_{2 n} (\Z_2)$ of order
\beq
| {\rm O}^-_{2 n} (\Z_2) | = 2 \cdot 2^{n (n - 1)} (2^n + 1) \prod_{i = 1}^{n - 1} (2^{2 i} - 1) ,
\eeq
so the cardinality of the orbit is
\beq
\frac{| \Sp_{2 n} (\Z_2) |}{| {\rm O}^-_{2 n} (\Z_2) |} = 2^{2 n - 1} - 2^{n -1} .
\eeq

The sum of the cardinalities of the two orbits of course equal the cardinality $2^{2 n}$ of the set of all quadratic forms obeying (\ref{curvature}). But for a given manifold $M$, not all such quadratic forms need to appear for any spin structure $s \in \cS$ on $M$. (Obviously, at most $2^{b_1 (M)}$ different quadratic forms can appear, which gives a restriction for a manifold $M$ such that $b_1 (M) \leq b_3 (M)$.) I am not aware of any example from the second orbit, and I conjecture that in fact only the first orbit can appear. A proof of this (or a counterexample) would be interesting. In any case, for the remainder of this paper, we will only consider the first orbit, which certainly leads to the nicest looking formulas.

In the next section, we will construct a basis for the vector space $V \simeq H^0 (T, \cL^N)$ that depends on a lifting of $J \in \cJ$ to the universal covering space $\overline{\cJ}$. This defines a global holomorphic frame for a holomorphic vector bundle over the topologically trivial space $\overline{\cJ}$. (This bundle is of course trivial.) The $\Sp_{2 n} (\Z)$ subgroup of the automorphism group of $H^3 (M, \Z)$ that stabilizes $\omega$ acts on $\overline{\cJ}$. It follows from the above that the subgroup $\Lambda_{(\omega, u)}$ that also stabilizes a given quadratic form $u \in \cU$ (on the first orbit above) is isomorphic to the kernel of the composite homomorphism
\beq
p \colon \Sp_{2 n} (\Z) \rightarrow \Sp_{2 n} (\Z_2) \rightarrow \Sp_{2 n} (\Z_2) / {\rm O}^+_{2 n} (\Z_2) ,
\eeq
where the first map is reduction modulo two, and the second map is the quotient projection. Thus
\beq
\Lambda_{(\omega, u)} \simeq \ker p .
\eeq
In this way, we get a topologically non-trivial holomorphic vector bundle $\tilde{E}$ over the quotient space $\overline{\cJ} / \Lambda_{(\omega, u)}$. As described in the introduction, pullback by $\phi \colon \overline{\cG} / \Lambda_{(\sigma, s)} \rightarrow \overline{\cJ} / \Lambda_{(\omega, u)}$ defines the partition bundle $E$ over $\overline{\cG} / \Lambda_{(\sigma, s)}$ of which $Z$ is a section.

\section{The trivialization}
The universal cover $\overline{\cJ}$ of the space of flat K\"ahler metrics on $T$ can be identified with the genus $n$ Siegel upper half space, i.e. the space of complex, symmetric $n \times n$ matrices with positive definite imaginary part. Much of what follows is analogous to the theory of holomorphic modular forms for (a subgroup of) $\Sp_{2 n} (\Z)$ (see e.g. \cite{Mumford}), but since we prefer to work in a differential form notation instead of choosing a specific basis for $H^3 (M, \R)$, some formulas might look slightly unfamiliar.

We begin by defining the conjugate $S^*$ of a linear map $S$ from (a subspace of) $H^3 (M, \C)$ to (a subspace of) $H^3 (M, \C)$ by the requirement that
\beq
\int_M x \wedge S x^\prime = \int_M S^* x \wedge x^\prime
\eeq
for all applicable $x, x^\prime \in H^3 (M, \C)$. The complex structure on $T$ can then be described by a map
\beq
\tau \colon A \rightarrow B \otimes \C
\eeq
which is anti self-conjugate, i.e. 
\beq
\tau = - \tau^* , 
\eeq
and has positive definite imaginary part, i.e. 
\beq
\frac{1}{2 i} \int_M n \wedge (\tau - \bar{\tau}) n \geq 0
\eeq
for all $n \in A$. The intermediate Jacobian $T =  H^3 (M, \R) / H^3 (M, \Z)$ can then be identified as
\beq
T \simeq \frac{B \otimes \C}{B \oplus \tau A} .
\eeq

The space $V \simeq H^0 (T, \cL^N)$ can be identified with the space of holomorphic functions
\beq
\psi (\tau | \, . \, ) \colon  B \otimes \C \rightarrow \C 
\eeq
which satisfy the double quasi-periodicity conditions
\beq \label{quasi-periodicity}
\psi (\tau | z + m + \tau n) = \psi (\tau | z) \exp \left( - \frac{N}{2} n \wedge \tau n - N n \wedge z \right)
\eeq
for $z \in B \otimes \C$, $n \in A$, and $m \in B$. In this trivialization of $\cL^N$, the transition functions thus depend holomorphically on $z \in B \otimes \C$. Alternatively, we can work in a unitary trivialization of $\cL^N$, with the sections given by functions $\Psi (\tau, \bar{\tau} | \, . \, , \, . \,)$ whose transition functions are $U (1)$-valued: 
\bea
\Psi (\tau, \bar{\tau} | z + m + \tau n, \bar{z} + m + \bar{\tau} n) = \Psi (\tau, \bar{\tau} | z, \bar{z}) \;\;\;\;\;\; \cr
\times \exp \left( \frac{N}{2} n \wedge m + \frac{N}{2} (\tau - \bar{\tau})^{-1} (z - \bar{z}) \wedge m + \frac{N}{2} \left( \tau (\tau - \bar{\tau})^{-1} \bar{z} - \bar{\tau} (\tau - \bar{\tau})^{-1} z \right) \wedge n \right) . \cr
\eea
This makes it easy to verify the holonomies of $\cL^N$ along the curves defined by straight lines from $0$ to a point $\gammaÊ= n + m \in H^3 (M, \Z) = A \oplus B$: Putting $z = \bar{z} = 0$, we get
\beq
\Psi (\tau, \bar{\tau} | m + \tau n, m + \bar{\tau} n) = \Psi (\tau, \bar{\tau} | 0, 0) (-1)^{u (n + m)}
\eeq
with
\beq
(-1)^{u (n + m)} = \exp \left( \frac{N}{2} n \wedge m \right) . 
\eeq
In particular, the holonomies are indeed trivial for $\gamma \in A$ or $\gamma \in B$. The relationship between the two trivializations $\psi (\tau | \, . \, )$ and $\Psi (\tau, \bar{\tau} | \, . \, , \, . \,)$ is
\beq
\Psi (\tau, \bar{\tau} | z, \bar{z}) = \psi (\tau | z) \exp \left( \frac{N}{2} (\tau - \bar{\tau})^{-1} (z - \bar{z}) \wedge z \right) ,
\eeq
but henceforth, we will only use the holomorphic trivialization $\psi (\tau | \, . \, )$.

The hermitian inner product on $H^0 (T, \cL)$ is given by
\bea
\langle \psi | \psi^\prime \rangle & = &  \int_{T} d^n z \, d^n \bar{z} \overline{\Psi (\tau, \bar{\tau} | z, \bar{z})} \Psi^\prime (\tau, \bar{\tau} | z, \bar{z}) \cr
& = &  \int_{T} d^n z \, d^n \bar{z} \overline{\psi (\tau | z)} \psi^\prime (\tau | z) \exp \left( \frac{N}{2} (\tau - \bar{\tau})^{-1} (z - \bar{z}) \wedge (z - \bar{z}) \right) .
\eea

We will now construct a holomorphic frame $\{ \psi_{[a]} \}_{[a] \in \frac{1}{N} A / A}$ for $H^0 (T, \cL^N)$, characterized by its behavior under translations:
\beq \label{multipliers}
\psi_{[a]} (\tau | z + b^\prime + \tau a^\prime) =  \psi_{[a + a^\prime]} (\tau | z) \exp \left( -\frac{N}{2} a^\prime \wedge \tau a^\prime - N a^\prime \wedge z + N a \wedge b^\prime \right) 
\eeq
for $a^\prime \in \frac{1}{N} A$ and $b^\prime \in \frac{1}{N} B$. In particular, $\psi_{[a]}$ is an eigensection of $T_{b^\prime}$ with eigenvalue $\exp (N a \wedge b^\prime)$. Note that (\ref{quasi-periodicity}) follows from (\ref{multipliers}) by taking $a^\prime = n \in A$ and $b^\prime = m \in B$. These properties determine the $\psi_{[a]}$ uniquely, up to a common holomorphic $\tau$-dependent factor. A convenient choice is
\beq \label{frame}
\psi_{[a]} ( \tau | z) = \frac{1}{\theta (\tau | 0)} \sum_{n \in A} \exp \left( \frac{N}{2} (n + a) \wedge \tau (n + a) + N (n + a) \wedge z \right) .
\eeq
Here $\theta (\tau | 0)$ is the Riemann theta function evaluated at $z = 0$:
\beq
\theta (\tau | z) = \sum_{n \in A} \exp \left( n \wedge \tau n + n \wedge z \right) .
\eeq
(This is of course the unique holomorphic section of the bundle $\cL$.) \footnote{There is a large literature on the connection between $(2, 0)$ theory and theta functions, see e.g. \cite{Witten96, Henningson-Nilsson-Salomonson, Witten99, Moore-Witten, Bonelli}.}

With a decomposition $H^3 (M, \Z) = A \oplus B$ obeying (\ref{Lagrangian}) and (\ref{hyperbolic}), a general map $S \colon H^3 (M, \Z) \rightarrow H^3 (M, \Z)$ and its conjugate map $S^* \colon H^3 (M, \Z) \rightarrow H^3 (M, \Z)$ can be written as a matrices of maps
\beq
S = \left( \begin{matrix} \alpha & \beta \cr \gamma & \delta \end{matrix} \right) \colon \left( \begin{matrix} B \rightarrow B & A \rightarrow B \cr B \rightarrow A  & A \rightarrow A \end{matrix} \right) 
\eeq
and
\beq
S^* = \left( \begin{matrix} \delta^* & \beta^* \cr \gamma^* & \alpha^* \end{matrix} \right) \colon \left( \begin{matrix} B \rightarrow B & A \rightarrow B \cr B \rightarrow A  & A \rightarrow A \end{matrix} \right) .
\eeq
We say that $S$ is symplectic if it preserves the symplectic structure $\omega$, i.e. if
\beq
S^* S = S S^* = 1. 
\eeq
This amounts to the relations
\bea
\delta^* \alpha + \beta^* \gamma & = & 1 \cr
\delta^* \beta + \beta^* \delta & = & 0 \cr
\alpha^* \gamma + \gamma^* \alpha & = & 0
\eea
and
\bea
\alpha \delta^* + \beta \gamma^* & = & 1 \cr
\beta \alpha^*  + \alpha \beta^* & = & 0 \cr
\gamma \delta^* + \delta \gamma^* & = & 0 .
\eea
(The unfamiliar signs are due to the definition of conjugate maps with respect to the symplectic structure.) The group of such symplectic maps is isomorphic to $\Sp_{2 n} (\Z)$. 

A symplectic map $S$ acts on $\tau \colon A \rightarrow B \otimes \C$ and $z \in B \otimes \C$ according to
\bea
\tau & \mapsto & S \tau = (\alpha \tau + \beta)(\gamma \tau + \delta)^{-1} \cr
z & \mapsto & S z = (\gamma \tau + \delta)^{* -1} z .
\eea
It is useful to note that
\beq
S (b + \tau a) = \alpha b - \beta a + (S \tau) (-\gamma b + \delta a)
\eeq
for $b \in B \otimes \C$ and $a \in A \otimes \C$. If $\psi$ is a $\tau$-dependent family of holomorphic sections obeying (\ref{quasi-periodicity}) for all $\tau$, we can then define another such family $S \psi$ by
\beq
S \psi (\tau | z) = \psi \left(S \tau | S z \right) \exp \left( -\frac{N}{2} \gamma z \wedge S z \right) .
\eeq
To verify that $S \psi$ also obeys (\ref{quasi-periodicity}), we need to use the relations
\bea
\exp \left( \frac{N}{2} \delta n \wedge \beta n \right) & = & 1 \cr
\exp \left( \frac{N}{2} \gamma m \wedge \alpha m \right) & = & 1 \cr
\exp \left( N \gamma m \wedge \beta n \right) & = & 1 .
\eea
The last of these three conditions is clearly always satisfied, and the first two impose non-trivial restrictions on $S$ only for odd $N$. In fact, they then define the condition for the reduction of $S$ modulo two to lie in the subgroup $O^+_{2 n} (\Z_2)$ of $\Sp_{2 n} (\Z_2)$.

We now apply this construction to the functions $\psi_{[a]}$.  The new functions $S \psi_{[a]}$ obey the translation properties
\bea
S \psi_{[a]} (\tau | z + b^\prime + \tau a^\prime) & = & S \psi_{[a - \gamma b^\prime + \delta a^\prime]} (\tau | z) \cr
& & \times \exp \left( - \frac{N}{2} a^\prime \wedge \tau a^\prime - N a^\prime \wedge z + N a \wedge (\alpha b^\prime - \beta a^\prime) \right. \cr
& & \left. - \frac{N}{2} \delta a^\prime \wedge \beta a^\prime - N \beta a^\prime \wedge \gamma b^\prime - \frac{N}{2} \gamma b^\prime \wedge \alpha b^\prime \right) .
\eea

Since both $\{ \psi_{[a]} \}_{[a] \in \frac{1}{N} A / A}$ and $\{ S \psi_{[a]} \}_{[a] \in \frac{1}{N} A / A}$ are frames for $H^0 (T, \cL^N)$, there must be an invertible linear relationship between them. Indeed \cite{Henningson}, to obey (\ref{multipliers}), $\psi_{[a]} (\tau | z)$ must be given by a (possibly $\tau$-dependent) multiple of
\bea \label{transition_functions}
& & \frac{1}{N^n} \sum_{[b] \in \frac{1}{N} B / B} S \psi_{[0]} (\tau | z + b + \tau a) \exp \left( \frac{N}{2} a \wedge \tau a + N a \wedge z \right) \cr & & = \frac{1}{N^n} \sum_{[b] \in \frac{1}{N} B / B} S \psi_{[- \gamma b + \delta a]} (\tau | z) \times \exp \left( - \frac{N}{2} \delta a \wedge \beta a - N \beta a \wedge \gamma b - \frac{N}{2} \gamma b \wedge \alpha b \right) .
\eea
But with the choice of prefactor in (\ref{frame}), this multiple is in fact a constant given by an eight root of unity. (It depends on the transformation $S$, but not on $[a]$ or $[b]$, and is not given by any elementary expression. An analogous prefactor appears already in the transformation law of the Riemann theta function \cite{Mumford}.) This can be verified by checking the special cases when 
\beq
S = \left( \begin{matrix} \alpha & \beta \cr 0 & \delta \end{matrix} \right)
\eeq
or
\beq
S = \left( \begin{matrix} 0 & \beta \cr \gamma & 0 \end{matrix} \right) ,
\eeq
which generate the whole group of transformations. The partition bundle is thus related both to a holomorphic modular form of weight $1 / 2$ (the Riemann theta function) and a flat vector bundle (the constant transition functions in (\ref{transition_functions})).

\vspace*{5mm}
This research was supported by the G\"oran Gustafsson foundation and the Swedish Research Council.
 
\appendix
\section{Spin structures and quadratic forms for $M = T^6$}
We have
\beq
b_1 (T^6) = 6
\eeq
and
\beq
b_3 (T^6) = 2 n = 20 .
\eeq
The set $\cS$ of spin structures on $T^6$ is in a natural way isomorphic to the linear space $H^1 (T^6, \Z_2)$ over $\Z_2$ rather than being an affine space over it. Indeed, the $\SO (6)$ holonomies of a (flat) six torus are trivial, and a lifting of them to $\Spin (6)$ is described by stating whether they are given by a trivial or non-trivial element for the different cycles. The trivial spin structure $s_0 \in \cS$, for which all these liftings are trivial, then corresponds to the zero element $0 \in H^1 (T^6, \Z_2)$.

The $\SL_6 (\Z)$ mapping class group of $T^6$ acts by permutation on the set $\cS \simeq H^1 (T^6, \Z_2)$ of spin structures via its reduction modulo two $\SL_6 (\Z_2)$. The orbits are:
\beq
\begin{array}{lllr}
\mathrm{rank} & \mathrm{representative} & \mathrm{stabilizer} & \mathrm{cardinality} \cr
\hline
0 & 0 & \SL_6 (\Z_2) & 1 \cr
1 & e^1 & \SL_5 (\Z_2) \ltimes \Z_2^5 & \underline{63} \cr
& & & 2^6 = 64 \cr
\end{array} 
\eeq
Here $e^1, \ldots, e^6$ is a basis of $H^1 (T^6, \Z)$, and the order of the group $\SL_d (\Z_2)$ is
\beq
| \SL_d (\Z_2) | = 2^{d (d - 1) / 2} \prod_{i = 2}^d (2^i - 1) .
\eeq

For a given spin structure $s \in \cS$, we now wish to determine the corresponding quadratic form
\beq
u \colon H^3 (T^6, \Z_2) \rightarrow \Z_2 .
\eeq
Although it is not really necessary for our purposes, we start by listing the orbits of $\SL_6 (\Z_2)$ on $H^2 (T^6, \Z_2)$:
\beq
\begin{array}{lllr}
\mathrm{rank} & \mathrm{representative} & \mathrm{stabilizer} & \mathrm{cardinality} \cr
\hline
0 & 0 & \SL_6 (\Z_2) & 1 \cr
1 & e^1 e^2 & \SL_4 (\Z_2) \times \SL_2 (\Z_2) \ltimes \Z_2^8 & 651 \cr
2 & e^1 e^2 + e^3 e^4 & \Sp_4 (\Z_2) \times \SL_2 (\Z_2) \times \Z_2^8 & 18228 \cr
3 & e^1 e^2 + e^3 e^4 + e^5 e^6 & \Sp_6 (\Z_2) & \underline{13888} \cr
& & & 2^{15} = 32768 \cr
\end{array} 
\eeq
More to the point is to determine the orbits of $\SL_6 (\Z_2)$ on $H^3 (T^6, \Z_2)$:
\beq
\begin{array}{lllr}
\mathrm{rank} & \mathrm{representative} & \mathrm{stabilizer} & \mathrm{cardinality} \cr
\hline
0 & 0 & \SL_6 (\Z_2) & 1 \cr
1 & e^1 e^2 e^3 & \SL_3 (\Z_2) \times \SL_3 (\Z_2) \ltimes \Z_2^9 & 1395 \cr
2 & (e^1 e^2 + e^3 e^4) e^5 & \Sp_4 (\Z_2) \ltimes \Z_2^9 & 54684 \cr
2 & e^1 e^2 e^3 + e^4 e^5 e^6 & \SL_3 (\Z_2) \times \SL_3 (\Z_2) \ltimes \Z_2 & 357120 \cr
3 & e^1 e^2 e^6 + e^2 e^3 e^4 + e^1 e^3 e^5 & \SL_3 (\Z_2) \ltimes \Z_2^8 & 468720 \cr
4 & e^1 e^2 e^6 + e^2 e^3 e^4 + e^1 e^3 e^5 + e^4 e^5 e^6 & \mathrm{order\,} = 2^7 \cdot 3^3 \cdot 5 \cdot 7 & \underline{166656} \cr
& & & 2^{20} = 1048576 \cr \cr
\end{array}
\eeq
Beginning with the $\SL_6 (\Z_2)$ invariant trivial spin structure $s_0 = 0 \in H^1 (T^6, \Z_2)$, the values of the corresponding quadratic form $u_0$ when evaluated on $\gamma \in H^3 (T^6, \Z_2)$ can only depend on the orbit of $\gamma$ under $\SL_6 (\Z_2)$. According to (\ref{distribution}), we should have
\beq
u_0 (\gamma) = \left\{ \begin{array}{lll}
0 & \mathrm{\; for \;} 524800 & \mathrm{\; values \; of \;} \gamma \in H^3 (T^6, \Z_2) \cr
1 & \mathrm{\; for \;} 523776 & \mathrm{\; values \; of \;} \gamma \in H^3 (M, \Z_2) .
\end{array} \right.
\eeq
The only solution is
\beq
u_0 (\gamma) = \left\{ \begin{array}{lll} 
1 & \mathrm{for \;} \gamma \mathrm{\; in \; orbit \; of \;} e^1 e^2 e^3 + e^4 e^5 e^6 \cr 
1 & \mathrm{for \;} \gamma \mathrm{\; in \; orbit \; of \;} e^1 e^2 e^6 + e^2 e^3 e^4 + e^1 e^3 e^5 + e^4 e^5 e^6 \cr
0 & \mathrm{otherwise}. \end{array} \right.
\eeq
One can check that this is consistent with the condition (\ref{curvature}).
Continuing with a general spin structure $s \in \cS$, the difference $\delta = u - u_0$ between the corresponding quadratic form $u$ and the quadratic form $u_0$ pertaining to the trivial spin structure $s_0$ defines a {\it homomorphism}
\beq
\delta \colon H^3 (T^6, \Z_2) \rightarrow \Z_2 .
\eeq
By Poincar\`{e} duality, such a $\delta$ can be identified with an element of $H^3 (T^6, \Z_2)$. Clearly, $s = s_0$, i.e. $\delta = 0$, then corresponds to the element $0 \in H^3 (T^6, \Z_2)$. But since there is no $\SL_6 (\Z_2)$ orbit of cardinality $64 - 1 = 63$ on $H^3 (T^6, \Z_2)$, also $s \neq s_0$ must correspond to the element $0 \in H^3 (T^6, \Z_2)$, i.e. 
\beq
u = u_0
\eeq
for all $s \in \cS$. This independence of $u$ on $s \in \cS$ for the case of $M = T^6$ was first noted in \cite{Dolan-Nappi}, and was given a deeper explanation in \cite{Hopkins-Singer}.

\end{document}